\documentclass[aps,prl,twocolumn,groupedaddress,longbibliography]{revtex4-1}

\usepackage{graphicx}
\usepackage{amsmath}
\usepackage{mdwmath}

\begin{document}



\title{Characterization of Parity-Time Symmetry in Photonic Lattices Using Heesh-Shubnikov Group Theory}


\author{Adam Mock}
\email[]{mock1ap@cmich.edu}
\affiliation{School of Engineering and Technology, Central Michigan University, ET 100, Mount Pleasant, MI 48859, USA \\ and
Science of Advanced Materials Program Central Michigan University, Mount Pleasant, MI 48859, USA}


\date{\today}

\begin{abstract}

We investigate the properties of parity-time symmetric periodic photonic structures using Heesh-Shubnikov group theory.
Classical group theory cannot be used to categorize the symmetry of the eigenmodes because the time-inversion operator
is antiunitary.  Fortunately, corepresentations of Heesh-Shubnikov groups have been developed to characterize the effect of antiunitary
operators on eigenfunctions.  Using the example structure of a one-dimensional photonic lattice, we identify the corepresentations of eigenmodes at
both low and high symmetry points in the photonic band diagram.  We find that thresholdless parity-time transitions are associated
with particular classes of corepresentations.  The approach is completely general and can be applied to parity-time
symmetric photonic lattices of any dimension.  The predictive power of this approach provides a powerful design tool for parity-time symmetric
photonic device design.

\end{abstract}


\maketitle


Recently it has been shown that non-Hermitian Hamiltonians that are invariant under the combined operation of parity ($\mathcal{P}$) and time-inversion ($\mathcal{T}$)
possess either real eigenvalues or sets of paired complex conjugate eigenvalues~\cite{bender1998, bender1999, bender2002}.  
Whether an eigenstate of such a Hamiltonian has a real or complex
eigenvalue depends on (i) the precise spatial symmetry of the non-Hermitian potential and (ii) the degree of non-Hermiticity. 
In this study we focus on (i) and apply Heesh-Shubnikov~\cite{heesh1929,shubnikov1964} group theory to electromagnetic systems with $\mathcal{PT}$ symmetry to determine
which states are expected to have real or complex eigenvalues.  Because the conclusions are based entirely on symmetry and not
on the degree of non-Hermiticity as in (ii), we expect the eigenvalues to maintain their realness or complexity even in the limit of infinitesimal
non-Hermiticity.  Previously, the existence of complex conjugate eigenvalues with infinitesimal non-Hermiticity has been referred to as thresholdless
$\mathcal{PT}$ symmetry breaking.  However, because such a situation arises as a direct result of the particular symmetry of the Hamiltonian,  a more
accurate descriptor would be two-fold $\mathcal{PT}$-degeneracy ($n$-fold if more than two eigenmodes with complex conjugate eigenvalues are involved).
Note that these modes are not rigorously degenerate because only the real part of their eigenfrequencies are equal.
When the eigenvalue of an eigenstate changes from real to complex as a function of the non-Hermiticity factor (as in (ii)), then the $\mathcal{PT}$ symmetry has been broken.

\begin{figure}
\centering
\includegraphics[width=8.6cm]{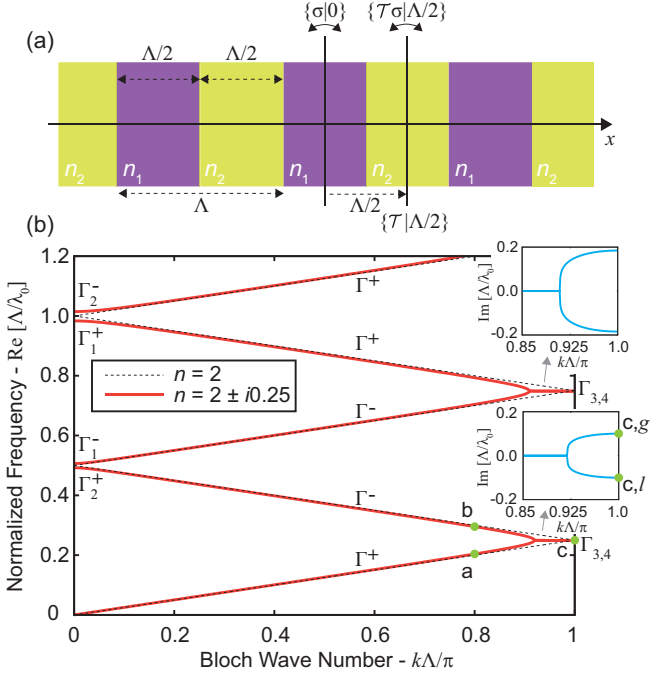}%
\caption{(a) Schematic diagram showing the one-dimensional $\mathcal{PT}$ symmetric photonic lattice.  Regions labeled $n_1$ provide gain ($n_1 = n_r + i n_i$),
and regions labeled $n_2$ provide loss ($n_2 = n_r - i n_i$) for positive $n_r$ and $n_i$.  Point and space group symmetry operations are labeled.  
(b) Photonic band diagram calculated using the plane wave expansion method.  Dashed black line: empty lattice band diagram.  Insets depict the imaginary
part of the frequency for modes with $\mathcal{PT}$-degeneracy.  $\lambda_0$ is the vacuum wavelength.  Corepresentation labels correspond to the
labels in Tables~\ref{tabC1h} and \ref{corep1} and Table VIII in~\cite{SI}. \label{fig_1dGeom} }
\end{figure}

Electromagnetics has proven a fruitful platform for exploring the consequences of $\mathcal{PT}$ symmetric 
Hamiltonians~\cite{elganainy2007, guo2009, mostafazadeh2009a, mostafazadeh2009b, ruter2010, longhi2010b, ctyroky2010, benisty2011,ge2012, hodeaei2014, feng2014, longhi2014, chang2014, peng2014, phang2015}. A $\mathcal{PT}$ symmetric electromagnetic
Hamiltonian can be created with appropriate spatial arrangements of regions in which electromagnetic waves experience gain or loss.
The gain and loss appear in the time-harmonic Maxwell equations as a complex index of refraction $n=n_r \pm i n_i$ ($+$ for gain, $-$ for loss), and the imaginary part $n_i$
is the non-Hermiticity factor.
Recent studies have shown that the modes of spatially periodic structures with $\mathcal{PT}$ symmetry exhibit a wide variety of behavior that depends on their
location on a band diagram: modes can be non-degenerate, ``classically degenerate'' or $\mathcal{PT}$-degenerate, and the $\mathcal{PT}$-degeneracy can be thresholdless
or be a function of a non-Hermiticity 
factor~\cite{musslimani2008, makris2008, zhou2010, lin2011,  szameit2011, regensburger2012, feng2013, regensburger2014, alaeian2014, zhu2014, xie2014, yannopapas2014, kulishov2014, zhu2015, ge2015, wang2015,  agarwal2015,  ge2015b, ding2015, mock2016, cerjan2016}.
Presently we investigate the $\mathcal{PT}$ symmetry classification of modes in a one-dimensional (1D) $\mathcal{PT}$ symmetric 
photonic lattice shown in Fig.~\ref{fig_1dGeom}(a).  The approach is completely general
and can be applied to $\mathcal{PT}$ symmetric geometries with periodicity in any dimension.  The general predictive power of the techniques presented here will help avoid
numerous unnecessary computations and provide valuable insight in $\mathcal{PT}$ symmetric photonic device design.

Heesh-Shubnikov groups~\cite{heesh1929, shubnikov1951, shubnikov1964} (also referred to as magnetic groups or 
color groups~\cite{cracknell1975a, cracknell1975b}) will be used to provide a general description of the role of symmetry
in determining whether eigenfunctions are expected to exhibit $\mathcal{PT}$-degeneracy with complex eigenfrequencies or are expected to be non-degenerate or classically degenerate
with real eigenfrequencies.
Heesh-Shubnikov groups describe the symmetry of regularly-shaped objects but whose components may have different colors.   Examples
include a square half of which is black and the other half is white or the \textit{taijitu} (yin and yang) symbol~\cite{cracknell1975b}.  
The development of Heesh-Shubnikov groups was motivated by studies of magnetic ordering
in ferromagnetic and ferroelectric materials~\cite{wigner1959, cracknell1975a, cracknell1975b}.  In these lattices the periodically arranged identical atoms are not distinguished by
color but, rather, by spin, and the same mathematical framework applies.

The electromagnetic wave equation in a source-free nonmagnetic medium in the frequency domain may be written as

\begin{equation}
\nabla \times [ \frac{1}{\epsilon(\vec r)} \nabla \times \vec H(\vec r) ] = \Xi \vec H(\vec r)  = \Big(\frac{\omega}{c} \Big)^2 \vec H(\vec r) 
\end{equation}

\noindent
where $c$ is the vacuum speed of light, and $\epsilon(\vec r)$ is the relative permittivity.  For $\mathcal{PT}$ symmetric systems in which 

\begin{equation}
[ \mathcal{PT}, \Xi ] = 0,
\end{equation}

\noindent
one has

\begin{eqnarray}
\Xi \mathcal{PT} \vec H(\vec r)  = \mathcal{PT} \Xi \vec H(\vec r) & = &  \mathcal{PT} \Big(\frac{\omega}{c} \Big)^2 \vec H(\vec r) \nonumber \\
& = &  \Big(\frac{\omega^{*}}{c} \Big)^2 \mathcal{PT} \vec H(\vec r).
\label{ccEigF}
\end{eqnarray}

\noindent
So if $\vec H(\vec r) $ is an eigenfunction with frequency $\omega$, then $\mathcal{PT} \vec H(\vec r)$ is also an eigenfunction but with frequency $\omega^*$. 
The complex conjugation of the eigenvalue is typical of antilinear operators.  $\mathcal{PT}$ is also an antiunitary operation which excludes the application
of representation theory based on classical groups.  Rather, corepresentation theory based on Heesh-Shubnikov groups is required.

The symmetry elements of the 1D periodic $\mathcal{PT}$ symmetric structure are shown in Fig.~\ref{fig_1dGeom}(a) and are given 
by $e = \{E|0\}$, $m=\{\sigma|0\}$, $\xi = \{\mathcal{T}|a/2\}$,  $\mu = \{\mathcal{T}\sigma|a/2\}$.
Elements $e$ and $m$ are unitary operators, whereas $\xi$ and $\mu$ are antiunitary due to the presence of $\mathcal{T}$.
The breakdown of classical group theory when dealing with antiunitary operations is illustrated by considering the matrix representation of the operators.
Let $R_i$ denote the $i$th unitary operator ($e$ or $m$) and let $A_i$ denote the $i$th antiunitary operator ($\xi$ or $\mu$).  Let $\mathbf{\Gamma}(B)$ be the matrix representation of
unitary or antiunitary operator $B$.  Then the following classical conditions must hold for a valid group representation: 
$\mathbf{\Gamma}(R_i)\mathbf{\Gamma}(R_j) = \mathbf{\Gamma}(R_i R_j)$ and
$\mathbf{\Gamma}(R_i)\mathbf{\Gamma}(A_j) = \mathbf{\Gamma}(R_i A_j)$.
However, when an antiunitary representation occurs first on the left side, then the following conditions must hold
$\mathbf{\Gamma}(A_i)\mathbf{\Gamma}(R_j)^* = \mathbf{\Gamma}(A_i R_j)$ and
$\mathbf{\Gamma}(A_i)\mathbf{\Gamma}(A_j)^* = \mathbf{\Gamma}(A_i A_j)$.
The complex conjugation of the second term spawns the development of the non-classical Heesh-Shubnikov group corepresentation theory~\cite{wigner1959}.

Because $\xi \xi = \{E|a\}$ is a pure translation, the full space group must be employed.  Based on the Bloch form for modes of periodic systems, one can 
use a representation of the space group, $\exp(i k n a)$, where $n$ is an integer~\cite{heine1960, mock2010c}.  
Application of space groups is facilitated by identification of the \textit{little group} or group of $\vec k$ which consists of symmetry operations which send
$\vec k$ into $\vec k + \vec K$ where $\vec K$ is a reciprocal lattice vector~\cite{sakoda2001,tinkham1964}.  However, for \textit{Heesh-Shubnikov little groups} that 
include antiunitary operators, such a group includes (i) unitary elements of the space group that send $\vec k$ into $\vec k + \vec K$ (as before) and 
(ii) antiunitary elements of the space group that send $\vec k$ into $-\vec k + \vec K$~\cite{cracknell1975b}.

For a 1D lattice $\vec k = \hat{x} k$, so for brevity we will proceed with the scalar part $k$.  
In the following we consider the Heesh-Shubnikov little group (HSLG) representations at high symmetry points $k = 0$ and $k=\pi/\Lambda$ 
and at a low symmetry point in the first Brillouin zone ($0 < k < \pi/\Lambda$).
For $k=0$, the space group representation takes on only one value ($\exp(i 0 n a) =1$), and the HSLG includes all of the symmetry
operations $\mathcal{M}^{k=0} = (e, m, \xi, \mu)$.  This group is isomorphic to $C_{2v} (2mm)$ and the \textit{Vierergruppe}~\cite{tinkham1964, SI}.
The elements  $\mathcal{N} = (e, m)$ do not contain $\mathcal{T}$, and they form a unitary subgroup of index 2.  This subgroup is isomorphic to
$C_{1h} (m)$~\cite{tinkham1964, SI}.  The antiunitary elements form a coset of $\mathcal{N}$: $A\mathcal{N}$ for $A \in ( \xi, \mu)$.   
Therefore, this HSLG may be expressed as $\mathcal{M}^{k=0} = \mathcal{N} + A\mathcal{N} = C_{1h} + \{\mathcal{T} | \frac{\Lambda}{2} \} C_{1h}$.
The final equality uses $A = \xi$ and helps illustrate the structure of the group.  Ultimately the HSLG contains two $C_{1h}$ symmetry centers offset by $\Lambda/2$
and distinguished by complex conjugation $\mathcal{T}$.  Cracknell classifies Heesh-Shubnikov groups of this form as Type IV~\cite{cracknell1975a,cracknell1975b}.

Corepresentations of $\mathcal{M}$ fall into three categories~\cite{wigner1959,ElBatanouny2008}.  
To determine the category Dimmock and Wheeler~\cite{dimmock1962} devised a sum rule similar to a rule obtained earlier by Frobenius and Schur~\cite{frobenius1906}.
The Dimmock and Wheeler test is

\begin{equation}
  \sum_{B \in \mathcal{W}} \chi(B^2) = \begin{cases}
                                                                     n & \textrm{Type (a),}  \\
                                                                    -n & \textrm{Type (b),}    \\
                                                                    0   &  \textrm{Type (c),} 
  \end{cases} \label{dmT}
\end{equation}


\noindent
where $\chi(R)$ is the character of the classical representation of $R$, $n$ is the order of the unitary subgroup and $\mathcal{W}$ is the set of
antiunitary operators.  Type (a) corepresentations correspond to a single representation of the unitary subgroup, 
and no new degeneracy is introduced.  Type (b) corepresentations contain the same single representation of the unitary 
subgroup twice, and new degeneracy may appear.  Type (c) corepresentations contain two inequivalent correpresentations of the unitary subgroup, and new
degeneracy is introduced~\cite{wigner1959,ElBatanouny2008}.   The primary outcome of this work is that thresholdless $\mathcal{PT}$ transitions are associated with Type (b) and (c) 
corepresentations, and modes with real frequency eigenvalues have Type (a) corepresentations.

\begin{table}[ht]
\caption{Corepresentations of $\mathcal{M}^{k=0}$.}
\begin{center}
\begin{ruledtabular}
\begin{tabular}{c c | c c | c c }
Correp. &  $C_{1h} (m)$ 
                              & \hspace{0.2cm} $ e $ \hspace{0.2cm}   &  \hspace{0.2cm} $ m $ \hspace{0.2cm}   
                              & \hspace{0.2cm} $ \xi $ \hspace{0.2cm} &  \hspace{0.2cm} $ \mu $ \hspace{0.2cm}   \\

\hline
(a) & $A'$, $\mathbf{\Gamma}^+_1$  & 1  &   1  &  1   &    1     \\
(a) & $A'$, $\mathbf{\Gamma}^-_1$  & 1  &    1  &  -1   &  -1     \\
(a) & $A''$, $\mathbf{\Gamma}^+_2$ & 1  &  -1  &  1   &  -1    \\
(a) & $A''$, $\mathbf{\Gamma}^-_2$ & 1  &   -1  &  -1   &   1    \\

\end{tabular}
\end{ruledtabular}
\end{center}
\label{tabC1h}
\end{table}

To continue with the symmetry analysis at $k=0$, we perform the Dimmock and Wheeler test (Eq.~\ref{dmT}).  Squaring the antiunitary operators results in
$(\xi^2, \mu^2) = (e, e )$ which yields two Type (a) corepresentations since $\chi(e) = 1$~\cite{tinkham1964, SI}.  
The components of the $i$th Type (a) corepresentation $\mathbf{\Gamma}_i$ for the unitary elements $R \in \mathcal{N}$ are given by 
$\mathbf{\Gamma}_i(R) = \mathbf{\Delta}_i(R)$ where $\mathbf{\Delta}_i(R)$ is the $i$th classical representation of $R$ in $\mathcal{N}$.
The components of the $i$th Type (a) corepresentation $\mathbf{\Gamma}_i$ for the antiunitary elements $R \in \mathcal{W}$ are given by
$\mathbf{\Gamma}_i(RA) = \mathbf{\Delta}_i(R)  \boldsymbol{\beta}$ where $A \in \mathcal{W}$ is an arbitrary but fixed antiunitary operator and 
$\boldsymbol{\beta} \boldsymbol{\beta}^* = \mathbf{\Delta}_i(A^2)$~\cite{wigner1959, ElBatanouny2008}.  
Using $A = \xi$ results in $\boldsymbol{\beta} \boldsymbol{\beta}^* = \mathbf{\Delta}_i(e) = 1$, so $\beta = \exp(\pm i \theta)$ (boldface removed to indicate 
scalar for the 1D corepresentation) with real $\theta$, and the total number of corepresentations is doubled.
Table~\ref{tabC1h} summarizes the results using $\beta = \pm 1$.  
$A'$ and $A''$ label the classical representations of $C_{1h}$.  $\mathbf{\Gamma}^{\pm}_i$ labels the corepresentations
for $\beta = \pm 1$.  

\begin{figure}
\centering
\includegraphics[width=8.6cm]{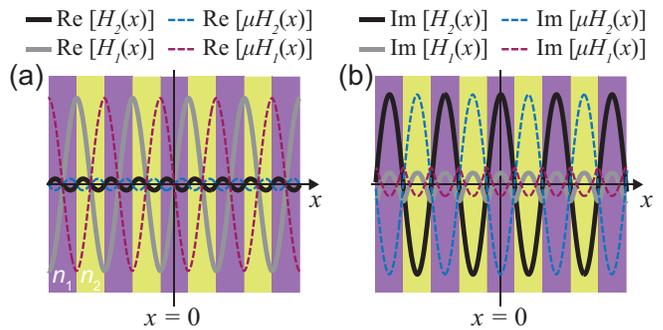}%
\caption{Magnetic field ($H_z(x)$) spatial distribution in $\mathcal{PT}$ symmetric 1D lattice at $k=0$.  $H_1$ corresponds to $\Gamma_1^-$ labeled
in Fig.~\ref{fig_1dGeom}(b), and $H_2$ corresponds to $\Gamma_2^+$.  Transformed fields are shown to verify the characters in Table~\ref{tabC1h}. \label{fieldsk0} }
\end{figure}

Because the corepresentations at $k = 0$ are all of Type (a), thresholdless $\mathcal{PT}$ degeneracy is not expected to occur there.
And because the classical representations of $C_{1h}$ are all 1D, classical degeneracy is also not expected at $k=0$.
The band diagram obtained using the plane wave expansion method~\cite{plihal1991, sakoda2001} and shown in Fig.~\ref{fig_1dGeom}(b) confirms this observation.  
Fig.~\ref{fieldsk0} illustrates the fields for points labeled $\Gamma_2^+$ and $\Gamma_1^-$ in Fig.~\ref{fig_1dGeom}(b)~\cite{SI}.
The character of the classical symmetry operations can be seen to be consistent with Table~\ref{tabC1h}.  
To illustrate the effect of an antiunitary operator, the result of operating with $\mu$ is shown; in both cases
operating with $\mu$ reproduces the same function but multiplied by $-1$ which is consistent with Table~\ref{tabC1h}.

\begin{table}[t]
\caption{Character table of $C_{2v}$ point group along with results of Dimmock and Wheeler test ($\alpha$) and corepresentation type (Correp.).}
\begin{center}
\begin{ruledtabular}
\begin{tabular}{c|c c c c | c c }
$C_{2v} (2mm)$ & \hspace{0.2cm} $ e $ \hspace{0.2cm} & \hspace{0.2cm}$ \overline{e}  $\hspace{0.2cm} 
                              & \hspace{0.2cm} $ m $ \hspace{0.2cm} & \hspace{0.2cm}$ \overline{m} $\hspace{0.2cm}  
                              & \hspace{0.2cm} $\alpha$ \hspace{0.2cm}  &  Correp. \\
\hline
$A_1$ & 1 & 1 & 1 & 1  &  4 &  (a)  \\
$A_2$ & 1 & 1 &-1 &-1 &  4 &  (a)  \\
$B_1$ & 1 &-1 & 1 &-1 &  0 &  (c)  \\
$B_2$ & 1 &-1 &-1 & 1 &  0 &  (c)  \\
\end{tabular}
\end{ruledtabular}
\end{center}
\label{tabC2v}
\end{table}

\begin{table*}[ht]
\caption{Corepresentations of $\mathcal{M}^{k=\pi / \Lambda}$.}
\begin{center}
\begin{ruledtabular}
\begin{tabular}{c c | c c c c | c c c c}
Correp. &  $C_{2v} (2mm)$ 
                              & \hspace{0.2cm} $ e $ \hspace{0.2cm} & \hspace{0.2cm}$ \overline{e}  $\hspace{0.2cm} 
                              & \hspace{0.2cm} $ m $ \hspace{0.2cm} & \hspace{0.2cm}$ \overline{m} $\hspace{0.2cm}  
                              & \hspace{0.2cm} $ \xi $ \hspace{0.2cm} & \hspace{0.2cm}$ \overline{\xi} $\hspace{0.2cm}  
                              & \hspace{0.2cm} $ \mu $ \hspace{0.2cm} & \hspace{0.2cm}$ \overline{\mu} $\hspace{0.2cm}  \\

\hline
(a) & $A_1$, $\mathbf{\Gamma}_1$ & 1 & 1 & 1 & 1  &  $(1)\beta$  &  $(1)\beta$  &  $(1)\beta$  &  $(1)\beta$  \\
(a) & $A_2$, $\mathbf{\Gamma}_2$ & 1 & 1 &-1 &-1 &  $(1)\beta$  &  $(1)\beta$  &  $(-1)\beta$  & $(-1)\beta$ \\
(c) & $B_1$, $\mathbf{\Gamma}_3$ &  

$\left( \begin{tabular}{cc}
         1 & 0  \\
         0 & 1  \\
        \end{tabular}
        \right)$
&
$\left( \begin{tabular}{cc}
         -1 & 0  \\
         0 & -1  \\
        \end{tabular}
        \right)$
& 
$\left( \begin{tabular}{cc}
         1 & 0  \\
         0 & -1  \\
        \end{tabular}
        \right)$
&
$\left( \begin{tabular}{cc}
         -1 & 0  \\
         0 & 1  \\
        \end{tabular}
        \right)$
&
$\left( \begin{tabular}{cc}
         0 & 1  \\
         -1 & 0  \\
        \end{tabular}
        \right)$
&
$\left( \begin{tabular}{cc}
         0 & -1  \\
         1 & 0  \\
        \end{tabular}
        \right)$
& 
$\left( \begin{tabular}{cc}
         0 & 1  \\
         1 & 0  \\
        \end{tabular}
        \right)$
&
$\left( \begin{tabular}{cc}
         0 & -1  \\
         -1 & 0  \\
        \end{tabular}
        \right)$
\\

(c) & $B_2$, $\mathbf{\Gamma}_4$ & 

$\left( \begin{tabular}{cc}
         1 & 0  \\
         0 & 1  \\
        \end{tabular}
        \right)$
&
$\left( \begin{tabular}{cc}
         -1 & 0  \\
         0 & -1  \\
        \end{tabular}
        \right)$
& 
$\left( \begin{tabular}{cc}
         -1 & 0  \\
         0 & 1  \\
        \end{tabular}
        \right)$
&
$\left( \begin{tabular}{cc}
         1 & 0  \\
         0 & -1  \\
        \end{tabular}
        \right)$
&
%
$\left( \begin{tabular}{cc}
         0 & 1  \\
         -1 & 0  \\
        \end{tabular}
        \right)$
&
$\left( \begin{tabular}{cc}
         0 & -1  \\
         1 & 0  \\
        \end{tabular}
        \right)$
& 
$\left( \begin{tabular}{cc}
         0 & -1  \\
         -1 & 0  \\
        \end{tabular}
        \right)$
&
$\left( \begin{tabular}{cc}
         0 & 1  \\
         1 & 0  \\
        \end{tabular}
        \right)$
\\

\end{tabular}
\end{ruledtabular}
\end{center}
\label{corep1}
\end{table*}

At $k=\pi/\Lambda$, the HSLG also includes all of the symmetry operations ($e, m, \xi, \mu$).
But the space group representation $\exp(i \frac{\pi}{\Lambda} n \Lambda) = \exp(i \pi n) = 1$ for $n$ even and $-1$ for $n$ odd.   To incorporate the
properties of the space group, the
symmetry elements are modified to $e = \{E|2n\Lambda\}$, $\overline{e} = \{E|2n\Lambda+\Lambda\}$, $m=\{\sigma | 2n\Lambda \}$, 
$\overline{m} =\{\sigma | 2n\Lambda + \Lambda \}$, $\xi = \{\mathcal{T}|2n\Lambda+\Lambda/2\}$,  
$\overline{\xi} = \{\mathcal{T}|2n\Lambda+\Lambda+\Lambda/2\}$, $\mu = \{\mathcal{T}\sigma|2n\Lambda + \Lambda/2 \}$ and 
$\overline{\mu} = \{\mathcal{T}\sigma|2n\Lambda + \Lambda + \Lambda/2\}$.  Therefore, the HSLG at $k=\pi/\Lambda$
is $\mathcal{M}^{k=\pi / \Lambda} = ( e, \overline{e}, m, \overline{m}, \xi, \overline{\xi}, \mu, \overline{\mu} )$ and is isomorphic to $C_{4v} (4mm)$~\cite{tinkham1964,SI}.  
The unitary subgroup of index 2 is $\mathcal{N} = ( e, \overline{e}, m, \overline{m} )$ and is isomorphic to $C_{2v} (2mm)$.
This HSLG can be expressed as $\mathcal{M}^{k=\pi / \Lambda} = C_{2v} + \{\mathcal{T} | \frac{\Lambda}{2} \} C_{2v}$.
The antiunitary elements are $\mathcal{W} = ( \xi, \overline{\xi}, \mu, \overline{\mu} )$. Squaring these elements results in
$( \xi^2, \overline{\xi}^2, \mu^2, \overline{\mu}^2 ) =  ( \overline{e},  \overline{e}, e, e )$.
Table~\ref{tabC2v} shows the character table for $C_{2v}$ and the result of the Dimmock and Wheeler test $ \alpha = \sum_{B \in \mathcal{W}} \chi(B^2) $.  
Representations $A_1$ and $A_2$ engender 
Type (a) corepresentations.  Physically, we seek corepresentations that change sign upon application of $e$ and $\overline{e}$.  Therefore we discard on physical
grounds the Type (a) corepresentations spawned by $A_1$ and $A_2$.  Further, assuming $A=\xi$, then $\boldsymbol{\beta} \boldsymbol{\beta}^* = \mathbf{\Delta}_i(\overline{e}) = -1$.  
That there is no solution for $\beta$ (boldface removed to indicate scalar for the 1D corepresentation) for these Type (a) 1D corepresentations is consistent with their unphysical nature.
However, for completeness, we show all of the corepresentations in Table~\ref{corep1}. 

Classical representations $B_1$ and $B_2$ engender Type (c) corepresentations.  The components of the $i$th Type (c) corepresentation $\mathbf{\Gamma}_i$ for the unitary elements 
 $R \in \mathcal{N}$ are given by~\cite{wigner1959, ElBatanouny2008}

\begin{equation}
\mathbf{\Gamma}_i(R) = \begin{pmatrix}
    \mathbf{\Delta}(R)  &  \mathbf{0} \\
    \mathbf{0}               &   \mathbf{\Delta}^*(S^{-1}RS) 
\end{pmatrix}
\end{equation}



\noindent
where $A = S\mathcal{T}$.  The components of the $i$th Type (c) correpresentation $\mathbf{\Gamma}_i$ for the antiunitary elements $R \in \mathcal{W}$ are given by

\begin{equation}
\mathbf{\Gamma}_i(R) = \begin{pmatrix}
   \mathbf{0}                    &  \mathbf{\Delta}^*(A^{-1}R) \\
   \mathbf{\Delta}(RA)   &  \mathbf{0}  
\end{pmatrix}.
\end{equation}


\noindent
Details of the calculation are provided in~\cite{SI}, and the results 
make up the last two rows of Table~\ref{corep1}~\cite{ElBatanouny2008}.  Corepresentations $\mathbf{\Gamma}_3$ and $\mathbf{\Gamma}_4$ are equivalent
because they can be transformed into each other via $U \mathbf{\Gamma}_3 U^{-1} = \mathbf{\Gamma}_4$ for the unitary elements and 
$U \mathbf{\Gamma}_3 (U^*)^{-1} = \mathbf{\Gamma}_4$ for the antiunitary elements~\cite{wigner1959,ElBatanouny2008,SI}.   

Because the $\mathbf{\Gamma}_{3,4}$ corepresentation changes sign between $e$ and $\overline{e}$, only this corepresentation is physically valid.
Therefore, every $k = \pi / \Lambda$ eigenstate of the $\mathcal{PT}$ symmetric 1D photonic lattice in Fig.~\ref{fig_1dGeom}(a) belongs to 
a two-dimensional (2D) Type (c) corepresentation.  
The photonic band structure displayed in Fig.~\ref{fig_1dGeom}(b) shows that 
coupled modes with complex conjugate eigenfrequencies form at every empty-lattice band crossing that occurs at $k = \pi / \Lambda$~\cite{SI}.
Since every mode at $k = \pi / \Lambda$ exhibits a thresholdless $\mathcal{PT}$ transition, we conclude that Type (c) 2D corepresentations are associated with 
thresholdless $\mathcal{PT}$-degeneracy.

\begin{figure}
\centering
\includegraphics[width=8.6cm]{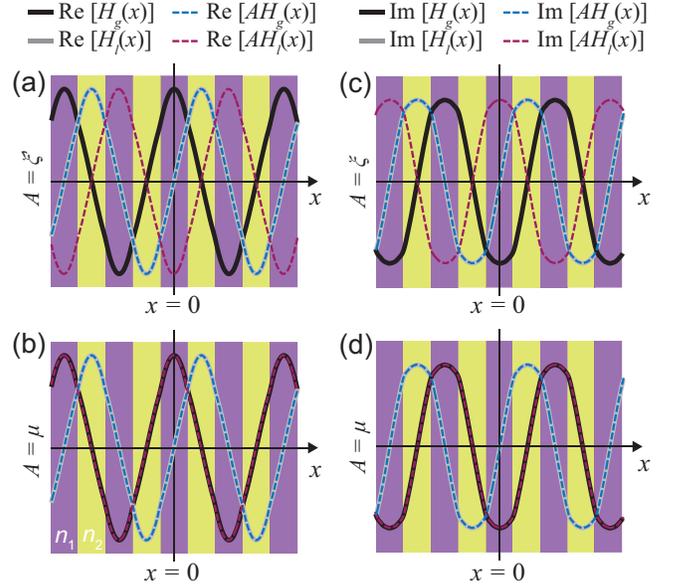}%
\caption{Magnetic field ($H_z(x)$) spatial distribution in $\mathcal{PT}$ symmetric 1D lattice at $k=\pi / \Lambda$ (point c in Fig.~\ref{fig_1dGeom}(b)).  
$H_g$ corresponds to gain mode labeled c,g in Fig.~\ref{fig_1dGeom}(b), and $H_l$ corresponds to the loss mode labeled c,l.  
Transformed fields are shown to verify the characters in Table~\ref{corep1}.  \label{fieldskpi} }
\end{figure}

In the $\mathcal{PT}$-degenerate
regime, two coupled eigenstates have complex conjugate eigenfrequencies.  Assuming a time-reference of $\exp(-\i \omega t)$, the mode with positive imaginary
frequency is the ``gain mode'', and the mode with negative imaginary frequency is the ``loss mode''. 
Figure~\ref{fieldskpi} illustrates the spatial field distribution for the two modes at $k=\pi/\Lambda$ at the frequency
$\Lambda/\lambda_0 \approx 0.25\pm i0.1$ (indicated by the green dots labeled `c', `c,g' and `c,l' in Fig.~\ref{fig_1dGeom}(b)).  
That these eigenfunctions possess the symmetry properties of the matrix corepresentations of the unitary operators shown in Table~\ref{corep1} is clear by inspection.  To confirm that
these modes also possess the symmetry properties of the matrix corepresentations of the antiunitary operators, $\xi$ and $\mu$ were applied to the gain and loss
eigenfunctions.  The transformed eigenfunction is represented by a dashed line.  As predicted, the gain mode transforms into the loss mode for $\xi$ and $\mu$,
and the loss mode transforms into the gain mode for $\mu$ and into its negative for $\xi$~\cite{SI}.

Finally, consider a wave vector at a low symmetry position in the first Brillouin zone (i.e. $k\neq 0, \pi/\Lambda$).  
For definiteness, we take $k = 0.8(\pi / \Lambda)$.  In this case
the only unitary operation that takes $\vec k$ to $\vec k + \vec K$ is the identity $E$, and the only antiunitary operation that takes $\vec k$ to $-\vec k + \vec K$ 
is $\mu$.  Because these symmetry operators do not result in pure translations, it is not necessary to employ the full space group.
Performing the Dimmock and Wheeler test results in
$\mu^2 = e$, so the corepresentations are all of Type (a), and no $\mathcal{PT}$-degeneracy is expected.  
The band diagram shown in Fig.~\ref{fig_1dGeom}(b) confirms this observation.
The corepresentation table and depiction of the fields for $k = 0.8(\pi / \Lambda)$ (labeled `a' and `b' in Fig.~\ref{fig_1dGeom}(b)) are provided in~\cite{SI}.

Application of Heesh-Shubnikov groups to a $\mathcal{PT}$ symmetric 1D lattice has allowed identification of points in the band diagram where thresholdless $\mathcal{PT}$-degeneracy
is expected.  Inspection of the band structure in Fig.~\ref{fig_1dGeom}(b) shows that there are 
$\mathcal{PT}$-degenerate modes for $k < \pi/\Lambda$.  This is not expected based on symmetry.  
As pointed out previously~\cite{mock2016}, the $\mathcal{PT}$ transition point shifts toward $k=0$ as $n_i$
is increased.  At the $\mathcal{PT}$ transition point, the modes with nominal Type (a) corepresentations transform into modes with Type (c) corepresentations.  
That this transition is a function of $n_i$, rather than symmetry, suggests that this phenomenon is indeed $\mathcal{PT}$ symmetry breaking.
As shown in~\cite{SI} increasing the non-Hermiticity factor to $n_i= 0.7$ can transform the third and fourth bands at $k=0$ from nominally Type (a) modes
to Type (c) modes.

The use of Heesh-Shubnikov group theory has facilitated the classification of modes of 1D photonic lattices that possess $\mathcal{PT}$ symmetry.  We found
points in the band structure in which thresholdless $\mathcal{PT}$-degeneracy occurs for every mode ($k = \pi / \Lambda$). Other than $k = \pi / \Lambda$, 
the modes are expected to be non-degenerate with real eigenvalues.  However, symmetry can be broken in these structures, and $\mathcal{PT}$ transitions are seen
for $k < \pi / \Lambda$ and depend of the non-Hermiticity factor $n_i$.  While a 1D lattice was the focus of this work, the approach is readily applicable to 2D and 3D
photonic crystals where the variety of modes is even richer.  Ultimately, we expect this analysis to be useful in the development of $\mathcal{PT}$ symmetric photonic
devices such as waveguides, cavities, delays and photonic crystal superprisms, to name a few.

\bibliography{myRef}
\end{document}


\begin{center}

\textbf{\textit{Supplemental Material For}} \\

\textbf{Characterization of Parity-Time Symmetry in Photonic Lattices Using Heesh-Shubnikov Group Theory} \\

Adam Mock\\
\textit{School of Engineering and Technology, Central Michigan University, ET 100, Mount Pleasant, MI 48859, USA \\ and
Science of Advanced Materials Program Central Michigan University, Mount Pleasant, MI 48859, USA}

\end{center}
%
%

\noindent
\textbf{1. Symmetry operators at $k=0$}

The group multiplication for the Heesh-Shubnikov little group (HSLG) of $k=0$, $\mathcal{M}^{k=0} = (e, m, \xi, \mu)$, is provided below.  Because
$k=0$, the space group representation does not play a role.  $\mathcal{M}^{k=0}$  is isomorphic to $C_{2v} (2mm)$.

\begin{table}[ht]
\caption{Multiplication table for $\mathcal{M}^{k=0}$.}
\begin{center}
\begin{tabular}{c | c | c | c | c |}

$C_{2v} (2mm)$ & \hspace{0.2cm} $ e $ \hspace{0.2cm} 
                              & \hspace{0.2cm} $ m $ \hspace{0.2cm} 
                              & \hspace{0.2cm} $ \xi $ \hspace{0.2cm} 
                              & \hspace{0.2cm} $ \mu $ \hspace{0.2cm}  \\
\hline
$ e $                       &  $ e $                       &  $ m $                       &  $ \xi $                         &  $ \mu $                      \\
\hline
$ m $                      &  $ m $                      &  $  e $                       & $ \mu$                       &  $ \xi    $                     \\
\hline 
$ \xi $                      &  $ \xi $                     &  $ \mu $                   &  $ e $                           &  $ m $                        \\
\hline
$ \mu $                   &  $ \mu $                    & $ \xi $                     &  $ m  $                        &  $  e  $                        \\
\hline
\end{tabular}
\end{center}
\label{tabMk0}
\end{table}

The unitary subgroup of $\mathcal{M}^{k=0}$ is isomorphic to $C_{1h} (m)$.

\begin{table}[ht]
\caption{Multiplication table for unitary subgroup of $\mathcal{M}^{k=0}$.}
\begin{center}
\begin{tabular}{c | c | c |}

$C_{1h} (m)$      & \hspace{0.2cm} $ e $ \hspace{0.2cm} 
                              & \hspace{0.2cm} $ m $ \hspace{0.2cm}   \\
\hline
$ e $                       &  $ e $                       &  $ m $                       \\
\hline
$ m $                      &  $ m $                      &  $  e $                        \\
\hline 
\end{tabular}
\end{center}
\label{tabusgMk0}
\end{table}

\begin{table}[ht]
\caption{Character table of $C_{1h} (m)$ point group along with results of Dimmock and Wheeler test ($\alpha$) and corepresentation type (Correp.).}
\begin{center}
\begin{tabular}{c | c c | c c }
\hline
\hline
$C_{1h} (m)$      & \hspace{0.2cm} $ e $ \hspace{0.2cm} 
                              & \hspace{0.2cm} $ m $ \hspace{0.2cm} 
                              & \hspace{0.2cm} $\alpha$ \hspace{0.2cm}  &  Correp. \\
\hline
$A'$  &  1 & 1 &  2 &  (a)  \\
$A''$ &  1 & -1 &  2 &  (a)  \\
\hline
\hline
\end{tabular}
\end{center}
\label{tabusgMk0C}
\end{table}

\noindent
\textbf{2. Discussion of modes at $k=0$}

To provide some insight into the lack of $\mathcal{PT}$ degeneracy at $k=0$, consider the effect of $\xi$
on an eigenfunction at $k = 0$: $\xi H^{0,i}(x) = [H^{0,i}(x-\Lambda/2)]^*$ where $H^{0,i}(x)$ is the eigenfunction of the $i$th band at $k = 0$.
From Table I in the main text we know that $\xi H^{0,i}(x) = \pm H^{0,i}(x)$, so we conclude $\pm H^{0,i}(x) = [H^{0,i}(x-\Lambda/2)]^*$.  The implication is that
the field intensity $|H^{0,i}(x)| = |[H^{0,i}(x-\Lambda/2)]^*| = |H^{0,i}(x-\Lambda/2)|$ is spatially periodic with a period of $\Lambda / 2$.  With this periodicity, spatially
shifting an eigenfunction by $\Lambda / 2$ reproduces the same field intensity instead of transforming into a field intensity with a different preferential overlap
with the gain or loss regions.  Therefore, these modes overlap the gain and loss regions equally which results in real eigenfrequencies.
The $\Lambda/2$ periodicity of $|H^{0,i}(x)|$ is apparent in the fields shown in Fig. 2 in the main text. \\

\noindent
\textbf{3. Symmetry operators at $k=\pi / \Lambda$}

At $k=\pi / \Lambda$, the space group representation plays a role in the analysis.
Space group operations are described by the Seitz operator $\{R|t \}$ which consists of a point symmetry operation $R$ followed by a translation $t$.
The multiplication operation between Seitz operators is given by

\[ \{R|t \} \{S|t' \} = \{RS|Rt'+t \} \]

As example consider $\overline{e}m = \{E|2m\Lambda+\Lambda\} \{\sigma | 2n\Lambda \} = \{ \sigma | 2(m+n)\Lambda + \Lambda \} \equiv \{ \sigma | 2n\Lambda + \Lambda \} = 
\overline{m} $ where we have noted
that a translation of $2(m+n)\Lambda$ is equivalent to $2n\Lambda$ for integer $m$ and $n$ at $k = \frac{\pi}{\Lambda}$.  Consider now 
$m \overline{e} =  \{\sigma | 2n\Lambda \}\{E|2m\Lambda + \Lambda \} = \{ \sigma | \sigma(2m\Lambda+\Lambda) + 2n\Lambda \} =  \{ \sigma | -(2m\Lambda + \Lambda) + 2n\Lambda \} 
= \{ \sigma | 2(n-m)\Lambda - \Lambda \} \equiv  \{ \sigma | 2n\Lambda+\Lambda \} = \overline{m} $.\\

When the Seitz operator contains the time inversion operation $\mathcal{T}$, it applies only to other time inversion operators and not to the point and spatial symmetry operations.
As example consider $m \mu = \{\sigma | 2m\Lambda \}  \{\mathcal{T}\sigma|2n\Lambda + \Lambda/2 \} =  \mathcal{T} \{\sigma \sigma | 2m\Lambda + \sigma(2n\Lambda + \Lambda/2) \} = 
\mathcal{T} \{E | 2m\Lambda - (2n\Lambda + \Lambda/2) \} = \{ \mathcal{T}  | 2(m-n) \Lambda - \Lambda/2) \} =  
\{ \mathcal{T}  | 2(m-n-1) \Lambda + \Lambda + \Lambda/2) \} \equiv  \{ \mathcal{T}  | 2n \Lambda + \Lambda + \Lambda/2) \}  = \overline{\xi}$. \\

Table~\ref{tabMkpi} shows the group multiplication table of the HSLG of $k = \frac{\pi}{\Lambda}$, $\mathcal{M}^{k = \pi / \Lambda}$.  $\mathcal{M}^{k = \pi / \Lambda}$ is isomorphic to
$C_{4v} (4mm)$.


\begin{table}[ht]
\caption{Multiplication table for $\mathcal{M}^{k=\pi / \Lambda}$.}
\begin{center}
\begin{tabular}{c | c | c | c | c | c | c | c | c |}

$C_{4v} (4mm)$ & \hspace{0.2cm} $ e $ \hspace{0.2cm} & \hspace{0.2cm} $ \overline{e} $ \hspace{0.2cm}
                              & \hspace{0.2cm} $ m $ \hspace{0.2cm} & \hspace{0.2cm} $ \overline{m} $ \hspace{0.2cm}
                              & \hspace{0.2cm} $ \xi $ \hspace{0.2cm} & \hspace{0.2cm} $ \overline{\xi} $ \hspace{0.2cm}
                              & \hspace{0.2cm} $ \mu $ \hspace{0.2cm} & \hspace{0.2cm} $ \overline{\mu} $ \hspace{0.2cm} \\
\hline
$ e $                       &  $ e$                       &  $ \overline{e} $   &  $ m $                        &  $ \overline{m} $   &  $ \xi $                       &  $ \overline{\xi} $   &  $ \mu $                     &  $ \overline{\mu} $ \\
\hline
$ \overline{e}  $    &  $ \overline{e} $    &  $  e $                     &  $ \overline{m}  $    &  $  m $                     &  $\overline{\xi}  $    &  $ \xi $                      &  $ \overline{\mu} $  &  $ \mu $ \\
\hline
$ m $                      &  $ m $                      & $ \overline{m}  $   &  $  e $                       &  $ \overline{e} $     & $ \overline{\mu} $  &  $ \mu $                   &  $ \overline{\xi}    $  &  $ \xi $ \\
\hline
$ \overline{m} $    &  $ \overline{m} $    &  $ m $                     &  $ \overline{e} $     &  $  e $                       &  $ \mu $                   &  $ \overline{\mu} $ &  $ \xi                      $  &  $ \overline{\xi}    $  \\
\hline
$ \xi $                      &  $ \xi $                     &  $ \overline{\xi} $  &  $ \mu $                   &  $ \overline{\mu} $  &  $ \overline{e} $     &  $  e $                       &  $ \overline{m} $    &  $ m  $ \\
\hline
$ \overline{\xi} $   & $ \overline{\xi} $    &  $ \xi $                     &  $ \overline{\mu} $ &  $ \mu $                   &  $ e $                        &  $ \overline{e} $     &  $ m $                       &  $ \overline{m} $ \\
\hline
$ \mu $                   &  $ \mu $                   & $ \overline{\mu} $  & $ \xi $                     &  $ \overline{\xi} $   &  $ m  $                       &  $ \overline{m} $    &  $  e $                       &   $ \overline{e} $ \\
\hline
$ \overline{\mu} $ & $ \overline{\mu} $ & $ \mu $                    &  $ \overline{\xi} $  & $ \xi $                       &  $ \overline{m} $     &  $ m  $                      &  $ \overline{e} $     &   $  e $ \\
\hline
\end{tabular}
\end{center}
\label{tabMkpi}
\end{table}

\begin{table}[ht]
\caption{Multiplication table for unitary subgroup of $\mathcal{M}^{k=\pi / \Lambda}$.  It is isomorphic to $C_{2v} (2mm)$.}
\begin{center}
\begin{tabular}{c | c | c | c | c |}

$C_{2v} (2mm)$ & \hspace{0.2cm} $ e                    $ \hspace{0.2cm} 
                              & \hspace{0.2cm} $ \overline{e}  $ \hspace{0.2cm} 
                              & \hspace{0.2cm} $ m                   $ \hspace{0.2cm} 
                              & \hspace{0.2cm} $ \overline{m} $ \hspace{0.2cm}  \\
\hline
$ e $                      &  $ e $                        &  $ \overline{e} $      &  $ m $                         &  $ \overline{m} $                      \\
\hline
$ \overline{e} $    &  $ \overline{e} $     &  $  e $                        & $ \overline{m} $        &  $ m $                     \\
\hline 
$ m $                     &  $ m $                       &  $ \overline{m} $     &  $ e $                           &  $ \overline{e} $                        \\
\hline
$ \overline{m} $   &  $ \overline{m} $    & $ m $                         &  $ \overline{e}  $       &  $  e  $                        \\
\hline
\end{tabular}
\end{center}
\label{tabusgMkpi}
\end{table}

\newpage

\noindent
\textbf{4. Construction of corepresentation table for $\mathcal{M}^{k=\pi / \Lambda}$} 

Here we provide calculation details for the construction of the corepresentation table for the group $\mathcal{M}^{k=\pi / \Lambda}$.
For Type (a) corepresentations, the elements of the unitary subgroup retain their classical representations $\mathbf{\Gamma}_i(R) = \mathbf{\Delta}_i(R)$ for $R \in \mathcal{N}$. \\

Corepresentations of the antiunitary elements are given in terms of the classical representations of the unitary subgroup according to 
$\mathbf{\Gamma}_i(RA) = \mathbf{\Delta}_i(R)\boldsymbol{\beta}$ with $\boldsymbol{\beta} \boldsymbol{\beta}^* = \mathbf{\Delta}_i(A^2)$.
Here we choose $A = \xi$.  Using a different operator for $A$ will give the same results.   \\

$\mathbf{\Gamma}_i(\xi e)                      = \mathbf{\Gamma}_i(\xi)                         = \mathbf{\Delta}_i(e)\boldsymbol{\beta}$  \\
$\mathbf{\Gamma}_i(\xi \overline{e})  = \mathbf{\Gamma}_i(\overline{\xi})     = \mathbf{\Delta}_i(\overline{e})\boldsymbol{\beta}$  \\
$\mathbf{\Gamma}_i(\xi m)                   = \mathbf{\Gamma}_i(\mu)                     =  \mathbf{\Delta}_i(m)\boldsymbol{\beta}$  \\
$\mathbf{\Gamma}_i(\xi \overline{m}) = \mathbf{\Gamma}_i(\overline{\mu})  = \mathbf{\Delta}_i(\overline{m})\boldsymbol{\beta}$  \\

The classical representations of the group $C_{2v}$ are provided in Table II in the main text.  The corepresentations resulting from this calculation 
make up the first two rows of Table III in the main text. \\

For Type (c) corepresentations,  the corepresentations of the elements of the unitary subgroup are given by Eq. 5 in the main text.  
With $A = \xi = S\mathcal{T}$, we identify $S = \{E|2n\Lambda + \Lambda/2 \}$ and $S^{-1} = \{E | -2n\Lambda - \Lambda/2 \}$.  
Determination of the matrix elements is shown in Table~\ref{tabSRS1}. 

\begin{table}[ht]
\caption{Determination of the matrix elements for the Type (c) corepresentations of the unitary operators in $\mathcal{M}^{k=\pi / \Lambda}$}
\begin{center}
\begin{tabular}{c | c | c}
$R$                      &      $RS$  &  $S^{-1}RS$ \\
\hline
$e$                      &      $\{E | 2n\Lambda + \Lambda/2 \}$  & $\{E | 0 \} = e$   \\
\hline
$\overline{e}$    &      $\{E | 2n\Lambda - \Lambda/2 \}$  & $\{E | -\Lambda \} = \overline{e}$   \\
\hline
$m $                    &      $\{\sigma | -\Lambda/2 \}$       &  $\{\sigma | -2n\Lambda - \Lambda \} = \overline{m}$\\
\hline
$\overline{m} $  &      $\{\sigma | \Lambda/2 \}$       &  $\{\sigma | -2n\Lambda  \} = m$ \\
\end{tabular}
\end{center}
\label{tabSRS1}
\end{table}

For Type (c) corepresentations,  the corepresentations of the antiunitary elements are given by Eq. 6 in the main text.  
With $A = \xi$, we identify $A^{-1} = \overline{\xi}$.  Determination of the matrix elements is shown in Table~\ref{tabBAAB1}. 

\begin{table}[ht]
\caption{Determination of the matrix elements for the Type (c) corepresentations of the antiunitary operators in $\mathcal{M}^{k=\pi / \Lambda}$.}
\begin{center}
\begin{tabular}{c | c | c }
$B$                          &      $BA                      $  & $ A^{-1}B              $  \\
\hline
$\xi$                        &      $ \overline{e}     $  & $ e                        $   \\
\hline
$\overline{\xi}$     &      $ e                        $   & $ \overline{e}     $   \\
\hline
$\mu $                    &      $  m                      $   &  $m                       $   \\
\hline
$\overline{\mu} $  &      $  \overline{m}  $   &  $\overline{m}     $ \\
\end{tabular}
\end{center}
\label{tabBAAB1}
\end{table}

The corepresentations resulting from this calculation make up the last two rows of Table III in the main text. \\

\noindent
\textbf{5. Unitary matrix that transforms $\mathbf{\Gamma_3}$ into $\mathbf{\Gamma_4}$ at $k=\pi / \Lambda$.}

Corepresentations $\mathbf{\Gamma}_3$ and $\mathbf{\Gamma}_4$ are equivalent
because they can be transformed into each other via $U \mathbf{\Gamma}_3 U^{-1} = \mathbf{\Gamma}_4$ for the unitary elements and 
$U \mathbf{\Gamma}_3 (U^*)^{-1} = \mathbf{\Gamma}_4$ for the antiunitary elements.  Any matrix of the form

\begin{displaymath}
U = \left( \begin{array}{cc}
                             0 & -e^{-i \theta} \\
                             e^{i \theta} & 0 \\
\end{array} \right)
\end{displaymath}

\noindent
accomplishes this transformation for real $\theta$. \\

\noindent
\textbf{6. Discussion of modes at $k=\pi / \Lambda$.}

Using the gain and loss modes ($H^{k,i}_g(x)$ and $H^{k,i}_l(x)$, 
respectively) as a basis of the corepresentation at wave number $k=\pi / \Lambda$ and empty-lattice band crossing point $i$,
Table III in the main text shows that when the unitary symmetry operators ($R$) are applied, the eigenfunctions transform according to diagonal matrices

\begin{displaymath}
R \left( \begin{array}{c}
                             H^{ k,i}_g(x)  \\
                             H^{ k,i}_l(x) \\
\end{array} \right) = 
\left( \begin{array}{cc}
                             r_{11} & 0           \\
                             0          & r_{22}  \\
\end{array} \right)
\left( \begin{array}{c}
                             H^{k,i}_g(x)  \\
                             H^{k,i}_l(x) \\
\end{array} \right) =
\left( \begin{array}{c}
                             r_{11} H^{ k,i}_g(x)  \\
                             r_{22} H^{ k,i}_l(x) \\
\end{array} \right)
\end{displaymath}

\noindent
showing that application of $R$ to the 2D basis does not mix or exchange the eigenfunctions.   This is to be contrasted to the transformation of the eigenfunctions
upon application of the antiunitary operators ($A$) which are represented by anti-symmetric matrices:

\begin{displaymath}
A \left( \begin{array}{c}
                             H^{k,i}_g(x)  \\
                             H^{k,i}_l(x) \\
\end{array} \right) = 
\left( \begin{array}{cc}
                             0            &   a_{12}   \\
                             a_{21}  &    0             \\
\end{array} \right)
\left( \begin{array}{c}
                             H^{k,i}_g(x)  \\
                             H^{k,i}_l(x) \\
\end{array} \right) =
\left( \begin{array}{c}
                             a_{12} H^{ k,i}_l(x)  \\
                             a_{21} H^{ k,i}_g(x) \\
\end{array} \right).
\end{displaymath}

\noindent
Here it is seen that application of the antiunitary operator transforms a loss mode \textit{into} a gain mode and vice versa.  As discussed previously [40]
gain and loss modes have similar symmetry, and if only the field intensity is visualized, the difference between them is a spatial shift such that the gain modes
preferentially overlap the gain regions, and the loss modes preferentially overlap the loss regions.  Because the antiunitary operators include
this spatial shift, the mode transformation properties of the matrix corepresentations make physical sense.  \\

\noindent
\textbf{7. Corepresentation for $\mathcal{M}^{k= 0.8 (\pi / \Lambda)}$}

Corepresentation for $\mathcal{M}^{k= 0.8 (\pi / \Lambda)}$ are Type (a).  
Using $A = \mu$ yields $\beta \beta^* = \mathbf{\Delta}_i(\mu^2) = 1$, so use $\beta = \pm1$. \\

$\mathbf{\Gamma}_i(R\mu) = \mathbf{\Gamma}_i(e \mu) = \mathbf{\Delta}_i(e)\beta = \beta = \pm 1$.

\begin{figure}
\centering
\includegraphics[width=12cm]{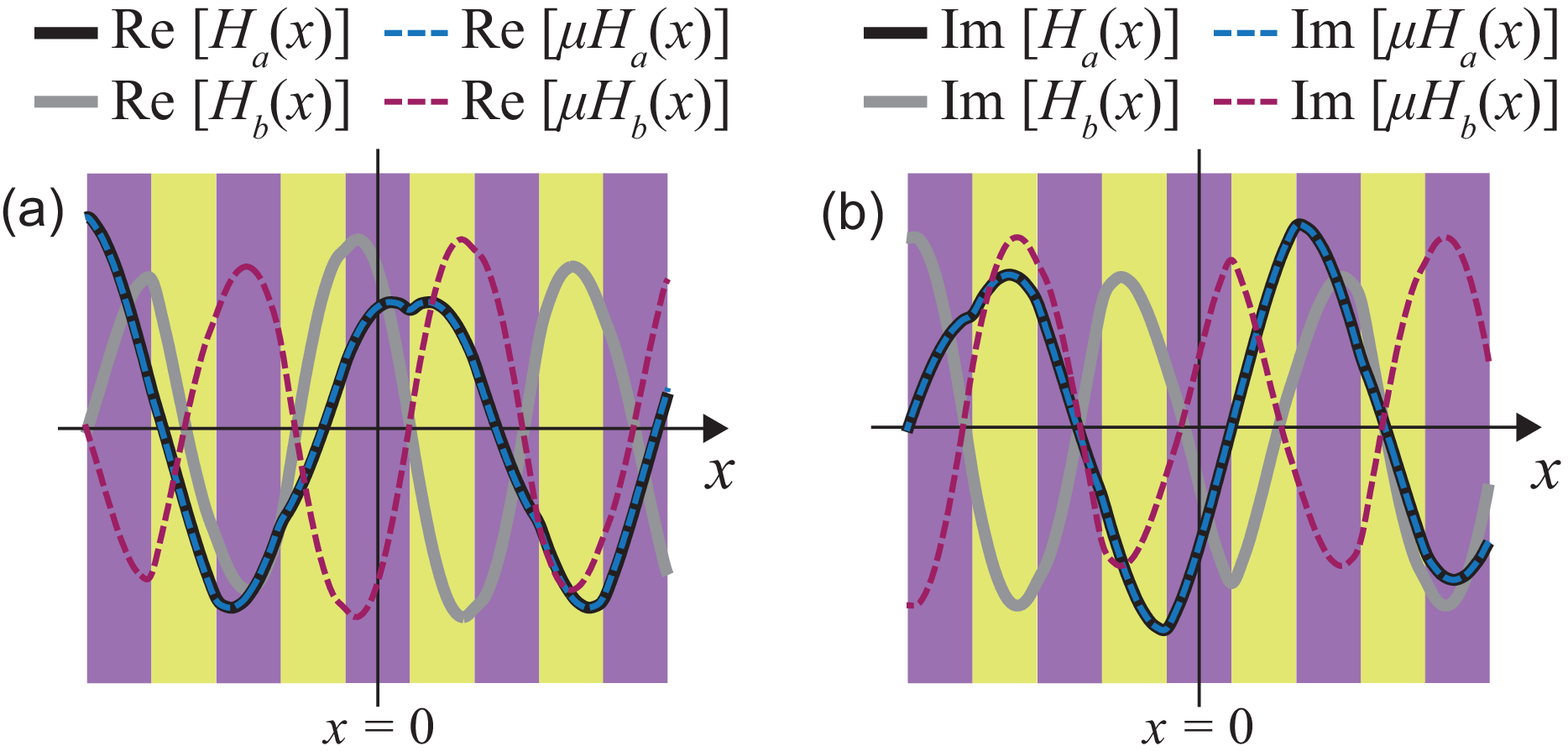}%
\caption{Magnetic field ($H_z(x)$) spatial distribution in $\mathcal{PT}$ symmetric 1D lattice at $k = 0.8 (\pi / \Lambda)$ (points a and b in Fig. 2(b) in the main text).  
Transformed fields are shown to verify the characters in Table~\ref{tabC1}.  \label{fieldsk0_8ni25}  }
\end{figure}

Inspection of the fields in Fig.~\ref{fieldsk0_8ni25} indicates that $H_a$ has corepresentation $\mathbf{\Gamma}^+$, and $H_b$ has corepresentation $\mathbf{\Gamma}^-$.

\begin{table}[ht]
\caption{Corepresentations of $\mathcal{M}^{k=0.8 (\pi / \Lambda)}$. }
\begin{center}
\begin{tabular}{c c | c | c }
\hline
\hline
Correp. &  $C_{1} (1)$ 
                              & \hspace{0.2cm} $ e $ \hspace{0.2cm}      
                              &  \hspace{0.2cm} $ \mu $ \hspace{0.2cm}   \\

\hline
(a) & $A$, $\mathbf{\Gamma}^+$  & 1  &   1      \\
(a) & $A$, $\mathbf{\Gamma}^-$  & 1   &  -1     \\
\hline
\hline
\end{tabular}
\end{center}
\label{tabC1}
\end{table}

\newpage

\noindent
\textbf{8. Broken $\mathcal{PT}$ symmetry}

\begin{figure}
\centering
\includegraphics[width=14cm]{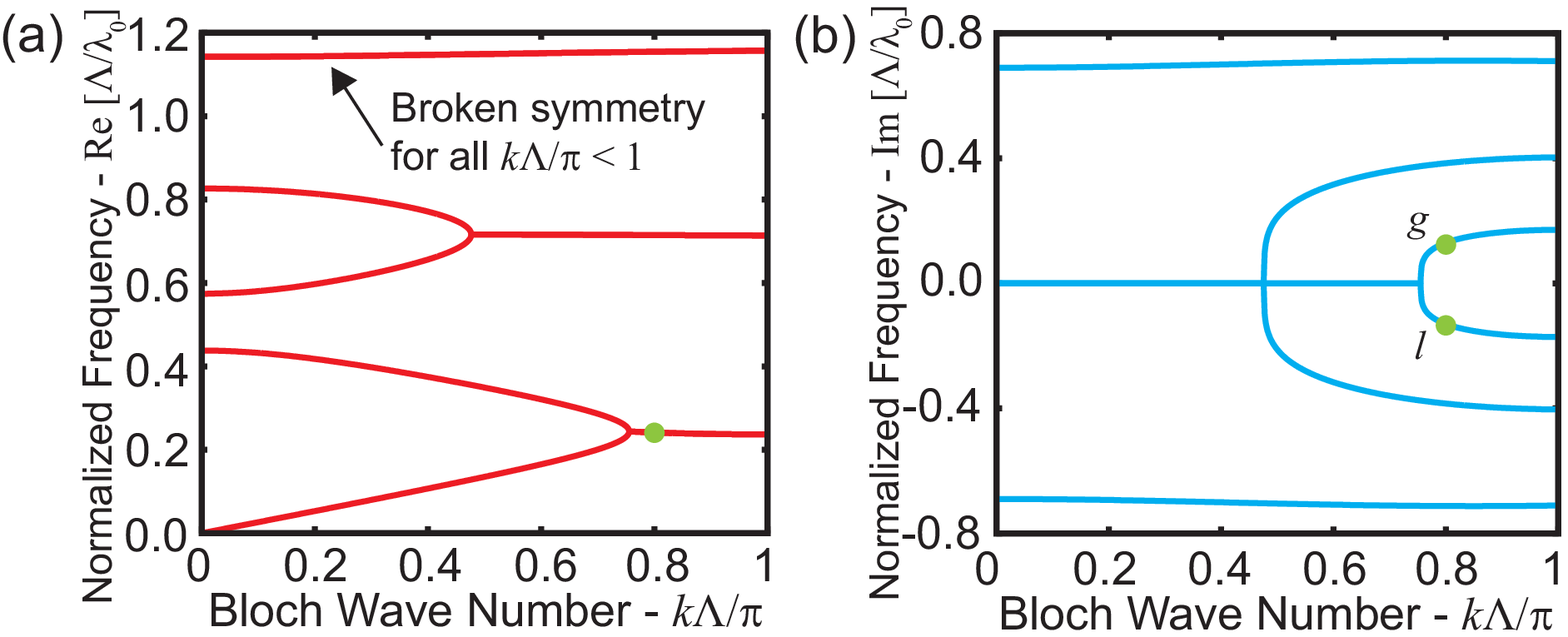}%
\caption{Photonic band diagram for the 1D photonic lattice shown in Fig. 1(a) in the main text with $n = 2 \pm i0.7$.  (a) Real part of the frequencies.  
(b) Imaginary part of the frequencies.  \label{bs70} }
\end{figure}

Fig.~\ref{bs70} shows a band diagram for the 1D photonic lattice shown in Fig. 1(a) in the main text but with $n = 2 \pm i0.7$.  This larger value for $n_i$ results
in $\mathcal{PT}$ transition points closer to $k=0$ than in Fig. 1(b).  In fact for the fifth and sixth bands, the $\mathcal{PT}$ transition point has reached $k=0$,
and these bands exhibit broken $\mathcal{PT}$ symmetry for $0 \leq k < \pi / \Lambda$.  \\

\begin{figure}
\centering
\includegraphics[width=12cm]{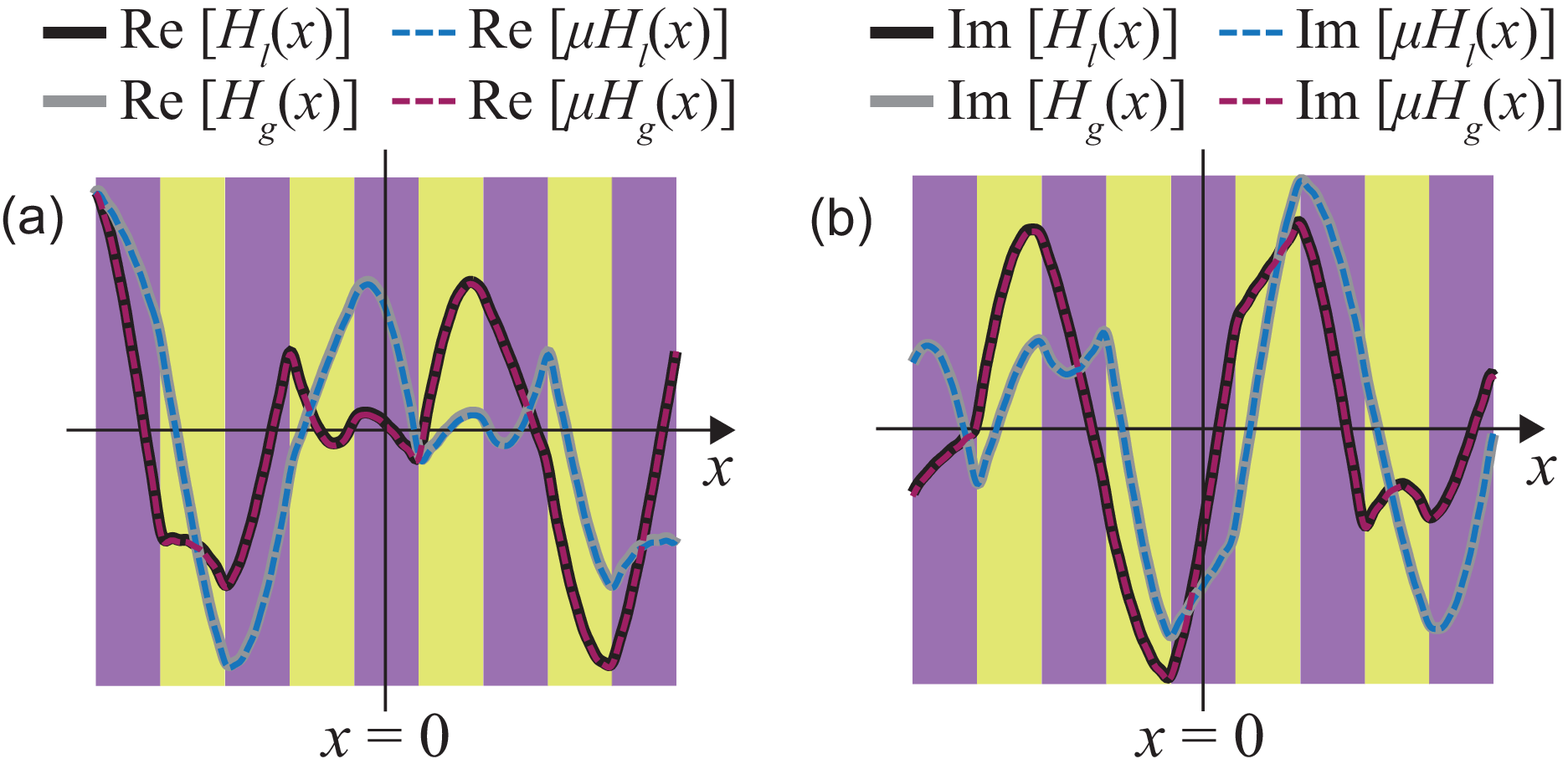}%
\caption{Magnetic field ($H_z(x)$) spatial distribution in $\mathcal{PT}$ symmetric 1D lattice with $n_i = 0.70$ at $k = 0.8 (\pi / \Lambda)$ (point marked by a circle in Fig.~\ref{bs70}).  
Transformed fields are shown to illustrate that mode symmetry differs from that predicted by group theory.   \label{fieldsk0_8ni70}  }
\end{figure}

Fig.~\ref{fieldsk0_8ni70} shows the spatial field distribution for the modes marked by a circle at $k=0.8(\pi / \Lambda)$ in Fig.~\ref{bs70}(a) and labeled g and l in Fig.~\ref{bs70}(b).
When $n_i = 0.25$, the fields at $k=0.8(\pi / \Lambda)$ are shown in Fig.~\ref{fieldsk0_8ni25}, and the characters in Table~\ref{tabC1} accurately
describe the symmetry of the fields.  When $n_i = 0.70$, the field symmetry is no longer of Type (a), and the mode is in the broken  $\mathcal{PT}$ symmetry regime. 
As shown in Fig.~\ref{fieldsk0_8ni70}, the fields
have Type (c) corepresentations where the antiunitary operator has a matrix corepresentation of the form 

\begin{displaymath}
\mu = \left( \begin{array}{cc}
                             0 & 1 \\
                            1 & 0 \\
\end{array} \right)
\end{displaymath}

This matrix is determined from observation of the field and does not follow from symmetry analysis.

